# The Smallsat Technology Accelerated Maturation Platform-1 (STAMP-1):

# A Proposal to Advance Ultraviolet Science, Workforce, and Technology for the Habitable Worlds Observatory

Kevin France[a], Jason Tumlinson[b,c], Brian Fleming[a], Mario Gennaro[b], Erika Hamden[d], Stephan R. McCandliss[c], Paul Scowen[e], Evgenya Shkolnik[f], Sarah Tuttle[g], Carlos J. Vargas[d], and Allison Youngblood[e]

[a]Laboratory for Atmospheric and Space Physics, University of Colorado, Boulder CO 80309, USA
[b]Space Telescope Science Institute, Baltimore MD 21218, USA
[c] Department of Physics and Astronomy, Johns Hopkins University, Baltimore, MD, 21218, USA
[d] Department of Astronomy & Steward Observatory, University of Arizona, Tucson, AZ, 85721, USA
[e]NASA/Goddard Space Flight Center, Greenbelt, MD, 20771, USA
[f]School of Earth and Space Exploration, Arizona State University, Tempe, AZ, 85287, USA
[g] Department of Astronomy, University of Washington, Seattle, WA, 98195, USA

## ABSTRACT

NASA's Great Observatories Maturation Program (GOMAP) will advance the science definition, technology, and workforce needed for the Habitable Worlds Observatory (HWO) with the goal of a Phase A start by the end of the current decade. GOMAP offers long-term cost and schedule savings compared to the 'TRL 6 by Preliminary Design Review' paradigm historically adopted by large NASA missions. Many of the key technologies in the development queue for HWO require the combined activities of 1) facility and process development for validation of technologies at the scale required for HWO and 2) deployment in the 'real world' environment of mission Integration & Test prior to on-orbit operations. We present a concept for the Smallsat Technology Accelerated Maturation Platform (STAMP), an integrated facility, laboratory, and instrument prototype development program that could be supported through the GOMAP framework and applied to any of NASA's Future Great Observatories (FGOs). This brief describes the recommendation for the first entrant into this program, "*STAMP-1*", an ESPA Grande-class mission advancing key technologies to enable the ultraviolet capabilities of HWO. STAMP-1 would advance new broadband optical coatings, high-sensitivity ultraviolet detector systems, and multi-object target selection technology to TRL 6 with a flight demonstration. *STAMP-1* advances HWO technology on an accelerated timescale, building on current ROSES SAT+APRA programs, reducing cost and schedule risk for HWO while conducting a compelling program of preparatory science and workforce development with direct benefits for HWO mission implementation in the 2030s.

kevin.france@colorado.edu

# 1. INTRODUCTION: TECHNOLOGY MATURATION FOR THE HABITABLE WORLDS OBSERVATORY

The Astro2020 Decadal Survey [1] prioritized a Large IR/optical/UV space telescope (later coined the "Habitable Worlds Observatory", HWO) to pursue an ambitious program of exoplanetary discovery and transformative cosmic origins astrophysics. The decadal survey report recognized that many of the galactic ecosystem, exoplanet, and stellar science goals of HWO require high-throughput imaging and spectroscopy at UV to optical wavelengths (100 – 1000 nm). The Astro2020 Panel on Electromagnetic Observations from Space 1 recommended that "... *[the] mission will also need focal plane instrumentation to acquire images and spectra over the range of 100 nm to 2 microns with parameters similar to cameras and spectrometers proposed for ... LUVOIR and HabEx.*" These instrument capabilities are required to enable compelling science across the whole scope of the survey.

The decadal survey also recognized that flagship astrophysics missions are multigenerational projects that present complex technical and management challenges. In the prominent case of JWST, these challenges caused numerous budget issues and programmatic delays [2]. The lessons learned from JWST informed the Astro2020 recommendation that NASA adopt an early program of technology maturation ahead of the pre-Phase A start of HWO to reduce cost and schedule risks to the mission. NASA is now implementing this recommendation in the form of the Great Observatories Maturation Program (GOMAP), a program that will refine the HWO science requirements from Astro2020 and prepare the requisite technologies for HWO prior to its formal new start later this decade.

Astro2020 highlighted the critical role that NASA's suborbital missions play in the advancement of critical-path technologies for small missions (NASEM 6-7,8). However, many of the key technologies in the queue for development for HWO are best demonstrated through the combined activities of instrument and observatory level integration and test (I&T), and long duration on-orbit use. The long-term performance and stability demonstration of advanced broadband optical coatings and ultraviolet-sensitive detectors can only be completely verified by their use as a system in the operational environment of space. Flight prototype demonstration fulfills the standard Technology Readiness Level (TRL) requirements, but goes beyond traditional metrics by providing 'real world' tests of the individual technologies as a system, presenting a true prototype-level demonstration of hardware as it will be used on HWO. This approach, long recognized as one of the motivations for the longevity of NASA's sounding rocket program, not only provides a path to TRL advancement, but provides real-time feedback to the technology community that is conducting the hands-on work of developing these components. Instead of waiting for HWO's instrument I&T in the mid-2030s to learn how this hardware works, we can learn these lessons a decade earlier and at a fraction of the cost.

## 1.1 Current Technology Maturation Avenues for Flagship Missions

Why are current funding and development efforts insufficient for accelerating technology development on the timescale and at the scope needed for HWO? As part of its recommendation for HWO, the Astro2020 decadal survey required that "*prior to commencing mission formulation, a successful [GOMAP] program must be completed, and a review held to assess plans in light of mission budgetary needs and fiscal realities.*" (NASEM 1-19) Key to this recommendation is that critical technology and infrastructure be matured (or have a path to maturation) ahead of the Phase A start to the mission, a recommendation also made by the LUVOIR concept study report (LUVOIR Final Report 2019).

The current paradigm, where technology is incrementally matured through a combination of Strategic Astrophysics Technology (SAT) and Astrophysics Research and Analysis (APRA) grants, is insufficient to achieve this maturation on the timescale needed for HWO. SAT awards are component-focused technology grants, in most cases designed around a limited number of formal lab-based maturation metrics that do not take into account how the hardware will be used as part of an integrated system. Furthermore, these programs often target metrics that do not derive from up-to-date mission science requirements. Because they are often not tied to specific science-driven capabilities, SAT and APRA have a disconnect with the groups defining scientific objectives and the systems-level hardware testing that reveals new challenges (and opportunities!) for the practical application of specific technologies.

APRA grants have led to key advancements in component-level hardware (see, e.g., the COPAG UV working group report; [3]), but these grants are too small to support scaling up of basic technology (and the Astro2020 decadal report notes *"basic technology grants are too small to support infrastructure"*) and rocket/balloon/cubesat missions have durations or volume allocations that are too short/small to demonstrate full prototype instruments in their relevant environments. Furthermore, APRA and SAT awards are financially insufficient to support the facility, technical, and workforce development required to rapidly advance the state-of-the-field to support a HWO gate review later this decade. The number of 3-year grant cycles required at the current level of SAT and APRA funding make completing HWO's technology maturation this decade through SAT and APRA alone impossible. Finally, mid-scale flight programs are not viable pathways to major technology advancement at the scope and timescale needed for HWO because they do not have technology as part of the program goals and therefore have risk averse technology postures (Pioneers, Explorer), long development timescales (Pioneers, Explorer), and/or sufficiently infrequent proposal opportunities (Explorers) that make these avenues infeasible for major technology advancement at the scope and timescale needed for HWO.

**1.2 Technology, Science and Workforce Advancement with the Smallsat Technology Accelerated Maturation Platform (STAMP)**

With these considerations in mind, we envision a new technology maturation concept that can be applied to all of NASA's Future Great Observatories (FGOs): the Smallsat Technology Accelerated Maturation Platform (STAMP). The STAMP is a dedicated laboratory development and flight demonstration program, where accelerated facility development advances technology that feeds a parallel small satellite demonstration mission. The lessons learned from the demonstration mission inform the ongoing technology and scaling efforts such that the final technology developed through GOMAP is ready to meet *all* the requirements of the full HWO mission, including how the hardware would be integrated and operate in its real-world setting. The STAMP combines the three pillars of GOMAP's charge: to advance science definition and technology for the FGOs while providing an inclusive mission framework that develops the science and engineering workforce that will lead the FGOs in the 2030s, 2040s, and beyond. We want to emphasize that the STAMP program is recommended *in addition to* the existing APRA and SAT program lines; APRA and SAT support the development of a range of technology at various maturation levels, and APRA supports novel science investigations and technology development for a balanced astrophysics portfolio. STAMP focuses on scaling up and systems-level validation of specific high-impact technologies, therefore, STAMP should be an augmentation of NASA's technology develop efforts to support the FGOs (ideally funded through the GOMAP program; we roughly estimate the cost of the STAMP technology and workforce maturation mission outlined below to have a budget of ~$50M, which can be compared to the anticipated $600M - $800M lifecycle GOMAP program cost for HWO) and not be implemented in place of existing development programs [4,5, Shkolnik in prep].

A more general description of the STAMP program and potential application to the full range of FGOs will be presented in a future work. This brief focuses on a concept for a first entrant into this program, the *STAMP-1* mission. As the initial prototype of this new technology development pathway, *STAMP-1* advances key technologies to enable the ultraviolet capabilities of HWO, including advanced broadband optical coatings, high-sensitivity ultraviolet detector systems, and multi-object selection technology (Section 2). The technology maturation mission would do this while carrying out a compelling program of preparatory science for HWO: a near- and far-ultraviolet (NUV and FUV) spectral imaging survey of nearby galaxies to identify the agents of galactic feedback and quantify the relationship between galactic outflows and the surrounding circumgalactic medium. This work addresses key topics in Astro2020's Cosmic Ecosystem theme (Section 2). In the process, *STAMP-1* would feature early-career researchers (ECRs) as deputies to every major science and instrument leadership position within the mission, and a science and instrument team that is comprised of more than 50% ECRs (Section 3). *STAMP-1* engages and trains the scientists and engineers who would lead the implementation and execution of HWO as the mission moves towards Phase A at the end of the current decade. Finally, STAMP-1 will include lessons about the overall viability of this development pathway for the X-ray and FIR flagships intended to follow HWO into GOMAP.

## 2. *STAMP-1*: TECHNOLOGY MATURATION, INSTRUMENT IMPLEMENTATION, AND DEMONSTRATION IN A COSMIC ECOSYSTEMS SCIENCE PROGRAM

HWO has three primary areas of technology development to carry out ahead of the mission confirmation review (expected prior to the end of the decade), as summarized in the TAG Technology Working Group technology gap list: 1) ultrastable telescope systems, 2) starlight suppression technology, and 3) components and observatory level development for high-efficiency UV imaging and spectroscopy. Ultrastable telescope systems are being developed under the ROSES D.19 element and startlight suppression is being advanced through SAT, Roman/CGI development, and EXEP investments. While the STAMP concept can provide on-orbit demonstration of telescope and high-contrast technologies, we focus here on STAMP-1 as a coherent, integrated approach to advancing the key UV technology needs for HWO. This section presents the high-priority ultraviolet hardware that would be matured under the *STAMP-1* program (Section 2.1), gives an overview of the demonstration instrument where they would be employed (Section 2.2), and presents a sample real-world HWO preparatory science investigation that would close the loop on the lifecycle demonstration of this hardware and pave the way for the development of HWO's ultraviolet instrument (Section 2.3).

**2.1 Technology Demonstration Goals**

For *STAMP-1*, there are three key enabling technologies, identified by the LUVOIR and HabEx study teams, reaffirmed by the HWO TAG, and subsequently prioritized in NASA's biennial technology gap list, to be matured:

1. Advanced broadband coatings: The telescope coatings for HWO are a focal point because their properties have an impact across the full mission, and any adopted technology must provide both UV sensitivity, high-broadband efficiency, and deposition uniformity to support both high-contrast imaging and science requiring access to ultraviolet wavelengths. *STAMP-1* provides a unique development platform to scale current state-of-the-art coatings to the size required for HWO, while meeting key demonstration milestones (Figure 1).
2. Large-format, photon-counting ultraviolet detector systems: HWO requires next-generation detector systems working at FUV and NUV wavelengths; *STAMP-1* demonstrates advanced MCP and UV-optimized CMOS flight arrays
3. Multi-object selection devices: multi-object spectroscopy for HWO requires selection devices beyond the MSA technology demonstrated on *JWST*. *STAMP-1* provides the 'real world' use of these next-generation devices in an orbital pathfinder instrument ahead of implementation on HWO.

As an example of the interplay between the laboratory and flight mission development aspects of the STAMP program, we consider the maturation of broadband mirror coatings (Figure 1). At present, advanced LiF-based protected aluminum optics offer the best combination of UV-through-NIR reflectance to achieve the full suite of HWO's science objectives, with hot deposition and chemically catalyzed deposition techniques offering as much as 20% higher reflectance per bounce in the FUV [13,14]. However, deposition facilities and process need to be scaled to the ~2-meter size required for HWO. Additionally, the uniformity and polarization properties need to be thoroughly characterized at this size scale, and optical coatings using the fabrication processes planned for HWO need to be tested in a long-term relevant environment. *STAMP-1* kicks off the engineering efforts needed to address all of these issues, which require investment and schedule acceleration beyond what can be achieved in the current ROSES SAT+APRA paradigm.

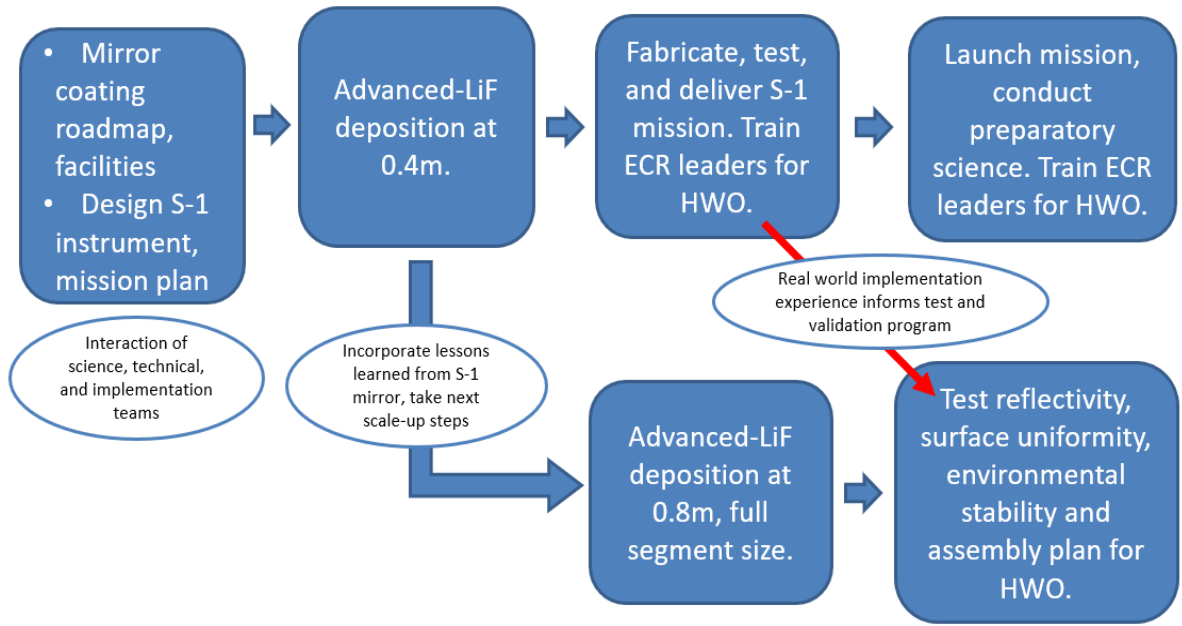

Figure 1 – Block diagram representation of the technology maturation path for HWO optical coatings in the STAMP framework. Unlike traditional technology maturation activities, STAMP provides direct interaction between the technology development and instrument implementation teams, providing process optimization feedback in nearly real time to mitigate cost and schedule risks associated with telescope development on HWO.

## 2.2 Prototype Instrument Implementation

The power of the *STAMP-1* instrument is in its multiplexing capability, a feature that derives directly from the combined advances in coatings (sensitivity to faint sources), high-sensitivity and low-background detectors (sensitivity to faint sources), and multi-object slitmasks (mapping large areas and different physical regions simultaneously). The *STAMP-1* design maps stellar populations (by observing FUV ions that are diagnostic of the effective temperature), mass outflow rates (by observing diagnostic FUV and NUV neutral and low-ionization tracers), and FUV emission lines from the circumgalactic halo (see Section 2.3).

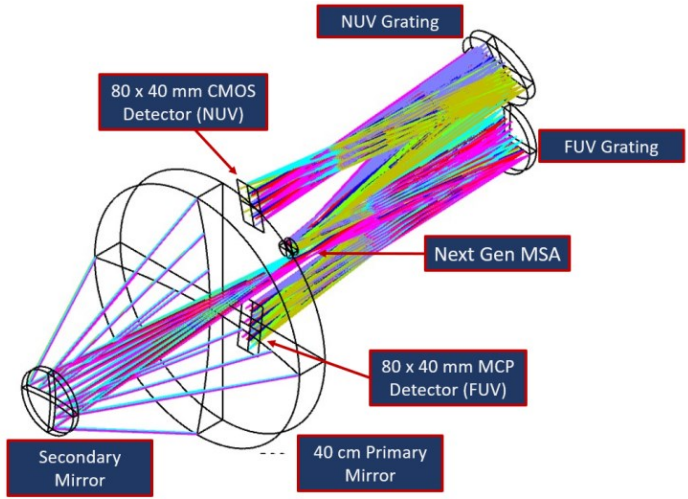

Figure 2 – Schematic raytrace of the STAMP-1 instrument. The 40 cm primary mirror feeds a two-channel spectrograph with a next-generation microshutter array as the multi-object selection mechanism at telescope focus.

The science payload employs the highest-priority enabling UV technologies for HWO to both advance their component-level TRL and demonstrate their application as an integrated system, providing a direct TRL and heritage link for their implementation on HWO (Figure 2). The telescope is a 40 cm diameter primary mirror Ritchey–Chrétien system. The light passes through a prototype 'next-generation' microshutter array (NG-MSA) device, which creates an electrostatically-actuated slit mask [11] that allows light from targets of interest into the spectrograph while excluding the remainder of the field. The prototype NG-MSA is a single wafer of 256 x 128 shutters (a unit cell of the 3 x 3 grid in development for HWO), where each shutter has a 100 μm x 200 μm pitch. Light that passes through the MSA continues on to a holographically ruled diffraction grating; FUV light is diffracted in the -1 order and directed to a photon-counting detector. The zero-order light off the FUV grating is directed to a second NUV grating that diffracts the light and directs it to the NUV focal plane array. Combining the area of the illuminated NG-MSA with the size of the FUV grating, the total field of view of the *STAMP-1* instrument is 32 x 32 arcminutes with each shutter subtending an angular size of approximately 9 x 18 arcseconds.

*STAMP-1* leverages recent advances in detector process development and readout electronics to provide long-duration flight demonstrations of large format, photon-counting detectors for both the FUV and NUV spectral regions, following the payload design originally recommended for LUVOIR [8]. The FUV detector is a large format (80 mm x 40 mm) microchannel plate (MCP) detector employing a cross-strip anode readout system. The cross-strip readout allows the MCP gain to be dropped by factors of ~10 relative to conventional delay line readouts, providing a correspondingly longer detector lifetime and builds on flight development for rocket missions. For the NUV focal plane, *STAMP-1* would employ an array of δ-doped CMOS detectors, including band-optimized anti-reflection coatings that suppress red response in silicon detectors. These devices leverage laboratory development through the APRA and SAT programs [12]. The peak effective areas of the STAMP-1 instrument are ~120 cm² ($\lambda_{peak}$ ~ 110nm) and ~70 cm² ($\lambda_{peak}$ ~ 280nm), respectively.

*Table 1 - The instrument parameter goals specified for the STAMP-1 instrument in the FUV (S1-FUV) and NUV (S1-NUV) channels. Both the spectral resolving power and angular resolution are limited by the ~5 arcsecond pointing stability anticipated for the small satellite attitude control system. The intrinsic instrument performance specifications are listed in parentheses in these categories.*

| *Instrument Parameter* | S1-FUV | S1-NUV |
|---|---|---|
| S/C-limited Spectral Resolving Power | 2,000 (4,000) | 2,500 (4,000) |
| Optimized Spectral Bandpass | 100 – 155nm (*restframe O VI – C IV*) | 240 – 300nm (*restframe Fe II – redshifted Mg II*) |
| S/C-limited Angular Resolution | 7 arcsec (3.5 arcsec) | 7 arcsec (3.7 arcsec) |
| Temporal Resolution | 1 msec | 1 sec |
| Field of View | 32′ × 32′ | |

Upon "authority to proceed", the facility development to process meter-class optics with the necessary overcoating would be developed (capability for Al+LiF deposition of optics as large as 1.5-m diameter at NASA/GSFC are being developed as part of an existing SAT). The optics demonstrated in this facility would proceed incrementally, moving from today's maximum of 20 cm to a 40 cm demonstration mirror. Following laboratory reflectance and uniformity testing, the 40 cm primary mirror (as well as secondary mirror and diffraction gratings) would be delivered to the *STAMP-1* instrument team where they would be integrated into the optomechanical structure of the payload ahead of instrument I&T. Mirrors developed in his program would also support testbed characterization of high-contrast imaging systems with HWO coating prescriptions. The full payload would undergo in-band performance and characterization testing, with witness samples tracked at every step to monitor coating stability. Any anomalies and lessons learned would be flowed back to the coating deposition team for process optimization when scaling up to the next milestone (80 cm) mirror and ultimately to a full HWO primary mirror segment demonstration (TBD pending TAG development). The interaction between the technology development team and the instrument development team provides the conduit to incorporate 'real world' lessons before the final HWO hardware is fabricated, reducing schedule and cost risk to the full HWO implementation.

Beyond the advanced broadband coatings, spacecraft, telescope, gratings, detectors, and prototype NG-MSA at the scale and specification for STAMP-1 are part of existing commercial or NASA-supported development efforts. We would expect telescope delivery within 12 months of order and deposition coating runs at this scale would begin immediately. Spacecraft, detectors, grating, and MSA would be expected between 18 – 21 months after receipt of funding, leaving 9 months for instrument assembly, calibration, and test and 6 months for integration with the spacecraft, observatory I&T, and delivery to the launch service provider. While we envision a 24-month science mission, all major technical milestones would be achieved with 6 months of in-flight operations.

## 2.3 Demonstration Science Program

Astro2020 highlighted three key science themes for Astronomy and Astrophysics in the 2020s and 2030s: Cosmic Ecosystems, Worlds and Suns in Context, and New Messengers and New Physics. A central element of Cosmic Ecosystems (Sections 1.1.3 and 2.3) is how "stellar feedback" drives the mass, energy, and chemical evolution of galaxies and the surrounding circumgalactic medium (CGM) with energetic radiation fields, stellar winds, and supernova explosions of the most massive stars. Understanding this feedback cycle was highlighted as a top priority for the Cosmic Ecosystems theme in the current decade, and was a primary driver for the multi-object capability in the LUVOIR and Habex studies. The decadal survey specifically called for close study of the relationship between star formation and the flows of matter in/out of nearby galaxies, where these processes can be observed "*in dramatic detail, revealing the full multiphase complexity of the local ecosystem*" (NASEM 1-8) and advance our understanding of how cool, warm, and hot gas is driven into the halos of galaxies, the properties of the gas within halos, and ultimately how that gas is recycled into future generations of stars.

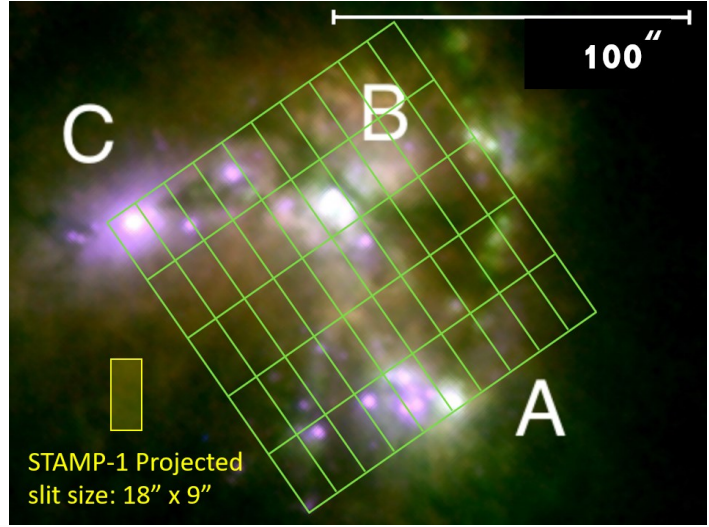

Figure 3 – Simulated overlay of the STAMP-1 MSA "footprint" on the central region of a nearby galaxy. Each shutter is represented as a yellow box, approximately 18" (cross-dispersion) x 9" (dispersion direction). This is an image of the starburst galaxy Haro 11, moved to a distance of 4.5 Mpc for this schematic figure (images from [6] and [7]); the overlay is meant to be illustrative only. This image highlights the ability of the STAMP-1 footprint to isolate various star-forming regions within a single galaxy but the full field-of-view of the STAMP-1 footprint is over 0.5°.

While advancing the technology and workforce that will enable UV imaging and spectroscopy on HWO, STAMP-1 will map the lifecycle of galaxy feedback, producing a dataset that will complement the high-spatial resolution local galaxy and higher-redshift halo studies to be carried out by HWO. STAMP-1 will use multi-object spectroscopy to simultaneously characterize the most massive stellar clusters with FUV photospheric and wind lines, measure the galactic outflows that drive gas into the circumgalactic medium through neutral and intermediate ionization FUV and NUV absorption lines, and trace that gas in emission in the CGM through FUV emission line mapping. The half-degree field-of-view of STAMP-1 enables this simultaneous mapping of nearby galaxies in a single observation, whereas HWO will likely have an angular footprint ~100 times smaller. This makes STAMP-1 an ideal tool to investigate the cycle of baryon feedback in the low-redshift universe where galaxies and galaxy halos are tens of arcminutes in angular extent. In addition, STAMP-1 will provide a precursor target list of particularly rich outflow and CGM sightlines for high sensitivity, high angular resolution follow-up with HWO. *STAMP-1* accomplishes this preparatory science program by implementing the first orbital multi-object spectrograph operating at FUV and NUV wavelengths, a pathfinder instrument for a future UV imaging and spectroscopic instrument on HWO.

While the specific scientific and technical requirements for HWO's instrument suite are still being refined by NASA's START, TAG, and nascent HWO project office, there is already a strong case for multi-object UV spectroscopy. This capability was highlighted as a driving requirement in Astro2020's Cosmic Ecosystems theme, it was recommended by the Astro2020 Electromagnetic Observations from Space-1 panel, and the HWO TAG has adopted the LUMOS-B instrument [8,9] as a baseline concept for HWO's UV capability. The *STAMP-1* instrument will provide multi-object spectroscopy at a spacecraft-limited angular resolution of approximately 7 arcseconds, enabling many tens of star-forming regions to be observed simultaneously (Figure 3). The spectral coverage and spectral resolution of the instrument 1) allow key mass and age diagnostics of the stellar population to be resolved (driving the short-wavelength FUV coverage, e.g., O VI 1032, 1038 Å, S IV 1063, 1073 Å, and P V 1118, 1128 Å) and 2) provide direct measurements of the outflows driven by the combined effects of stellar winds and supernovae in these star-forming regions. Strong outflow lines motivate FUV spectral coverage (Si II 1190, 1193, 1526, 1533 Å) and drive the NUV spectral bandpass (rest frame and low-redshift Fe II and Mg II; 2400 – 3000 Å).

The 40-cm telescope is sized to enable the survey goal of spectrally mapping 200 nearby galaxies in a 2-year primary science mission (noting that the technical mission would be achieved after 6 months). **The unique combination of multi-object spectroscopy and high-sensitivity coverage down to 100 nm is beyond the capability of Hubble or any other proposed mission.** We note that neither HST nor UVEX covers rest-frame O VI in an imaging spectroscopy mode. The bandpass and spectral resolution requirements of *STAMP-1* are driven by the need to cover the full set of stellar and gas diagnostics (FUV and NUV), with sufficient velocity resolution (~150 km/s) to resolve outflows from stellar clusters (typical outflow velocities range from 10s to ~1000 km/s; [10] and references therein); while demonstrating technology at the key wavelengths targeted by HWO.

*STAMP-1* simultaneously measures the ionization state and composition of the halo gas using low-resolution spectroscopy of wide angular fields in the halos of the target galaxies. Coverage of the most important cooling lines for gas between 10,000 – 300,000 K (O VI, Ly α, and C IV; rest wavelengths 1032 – 1550 Å, another driver of the FUV spectral coverage) over large angular extents with high sensitivity would allow STAMP-1 to map diffuse emission from circumgalactic halos in unprecedented detail and numbers (a key component of the "*Unveiling the Drivers of Galaxy Growth*" science priority area, NASEM 1-8). Taking into account telescope collecting area and realistic component efficiencies, this low-redshift galaxy survey (200 nearby galaxies) would be sensitive to ($> 8\sigma$) O VI surface brightness to ~$1 \times 10^{-18}$ erg cm$^{-2}$ s$^{-1}$ arcsec$^{-2}$ per 9′′ x 7′′ angular bin (this is the full slit width in the dispersion direction and the observatory angular resolution in the cross-dispersion axis) in a 200 ksec observation, and correspondingly lower surface brightness limits when binning over multiple shutters. This galaxy + halo survey would serve as a pathfinder scientific data set for galaxy-halo investigations over a wide range of redshifts and at much higher angular and spectral resolution with HWO. In addition, these observations will be made with an instrument similar in design and data format to the HWO UV instrument, providing a training ground to develop and analysis and data calibration techniques that will ultimately be needed for HWO.

### 3. WORKFORCE DEVELOPMENT IN THE STAMP FRAMEWORK

Astro2020 emphasized the lack of diversity at all levels of NASA mission leadership (NASEM Section 3; 6-10). NASA, and more specifically the GOMAP, has made inclusive workforce development a priority area for the pre-Phase A development of HWO. In the context of Explorer-class missions, Astro2020 noted: "*A first step to achieving a more diverse leadership pool is to broaden participation in technical, instrument, and leadership teams as a whole*" (NASEM 6-10); this initiative can extend to the science and technology preparatory efforts for HWO kickstarted by *STAMP-1*.

The organizational layout of the mission features 'deputies' in all key mission science roles (Figure 4), including deputy-PI (dPI), deputy-Project Scientist (dPS), deputy-Instrument Scientist (dIS), and in all the major engineering and program management positions (Program Manager, Project Systems Engineer, Instrument Systems Engineer, etc). These positions would be fully funded within the *STAMP-1* PI-managed budget and provide a direct route for the training of the scientists and engineers with the expertise to lead the future of HWO.

The *STAMP-1* science and instrument teams would be made up of > 50% early-career researchers (defined as less than 10 years from receipt of terminal degree); this builds the framework for critical relationships between observers, theorists, and instrumentalists that enables open communication and a cohesive development environment that can be applied to the forthcoming HWO science, instrument, and technology teams, while maintaining a connection to current space-flight mission expertise in science, engineering, and program management. The > 50% early-career researcher goal is achieved through pairing major mission leadership roles with early-career deputies (as implemented on the *Aspera* smallsat mission, [15]); the envisioned STAMP-1 mentoring approach emphasizes investment in hands-on mission experience for the group that will lead HWO in the 2030s and 2040s, while taking advantage of NASA's investments today. The STAMP program would recruit from a broad pool of early career scientists and engineers in a manner consistent with NASA Science Mission Directorate strategic priorities for Inclusion, Diversity, Equity, and Accessibility (IDEA)[1] and following the recommendations of the Astro2020 State of the Profession and Societal Impacts recommendations. Across the deputies, ~8 – 10 early career science team members, and ~8 – 10 early career engineers and program management personnel, we would imagine roughly two dozen early career scientists and engineers could train in key mission roles as part of an individual STAMP mission.

This emphasis on early-career development builds on the long-standing success of NASA's suborbital rocket and balloon programs ("*the program has demonstrated its efficacy in producing leaders for space missions*" NASEM Section 6.2.1.1), which has recently been adopted by cubesat and Pioneer missions such as *CUTE*, *SPARCS*, *SPRITE*, and *Aspera [16,17,18,15]*. It would be anticipated that some of the members of the STAMP-1 team would form part of an eventual science and engineering team for HWO's UV instrument – training people for this role is one of the goals of the STAMP program, – but the expectation for HWO is that all instruments will be solicited through open calls to the community; participation in a STAMP program is not a prerequisite for involvement in other HWO activities.

The early-career focus of the *STAMP-1* concept authors is balanced by mid-career leaders in ultraviolet science and instrumentation, leveraging their experience and NASA's investment in the institutional infrastructure and mission leadership to drive the preparation of HWO through the 2020s and into the Phase A and B activities in the 2030s. The STAMP early-career training and leadership mentoring framework is responsive to Astro2020s workforce development charge: "… *important that experienced PIs establish team roles that enable emerging leaders to gain experience*" (NSAEM 6-11). The primary mission management and systems engineering would be at NASA/GSFC to provide additional NASA guidance to the *STAMP-1* program and connection to the larger HWO development. The connection between the NASA management and *STAMP-1* science and instrument teams would develop the personal and institutional relationships that could be leveraged into the implementation of HWO. We would envision that the HWO project office at GSFC would also assign a 'GOMAP Project Scientist' to the *STAMP-1* project.

---

[1] https://science.nasa.gov/about-us/idea

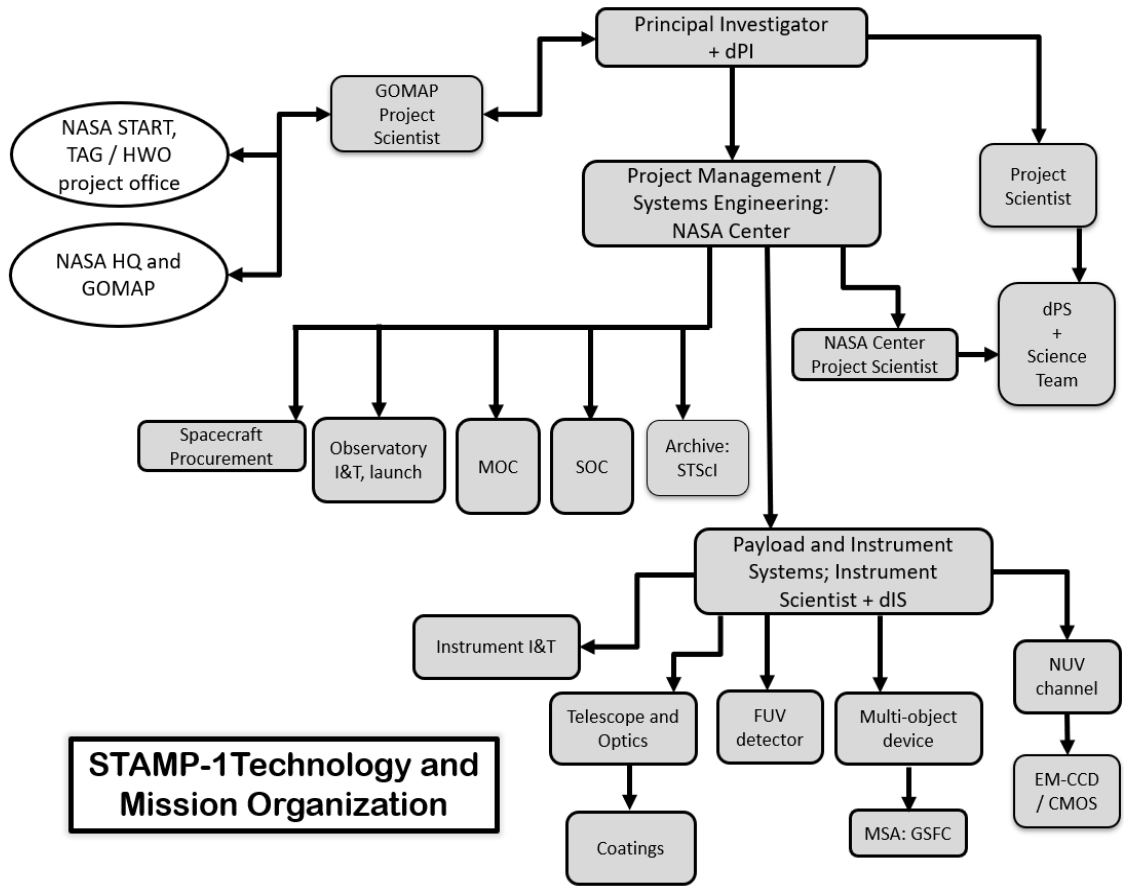

Figure 4 – Notional organizational block diagram for STAMP-1, leveraging existing NASA investments in institutional infrastructure and component level technology, while providing an inclusive federal/academic partnership that advances the development of early career scientists and engineers for leadership roles in the development and execution of HWO. Colors are representative of clustered expertise and investment across academia, federal, and non-profit research centers.

Finally, connection between the *STAMP-1* team, NASA HQ, the current START and TAG and/or the HWO project office is essential. These stakeholders would be regularly briefed on the progress of *STAMP-1* and the implication for the technology readiness of the ultraviolet hardware for HWO. These groups can then make informed instrument trade decisions based on current and projected performance through open and established lines of communication. To support this open connection between *STAMP-1* and the relevant stakeholders, the *STAMP-1* team supports a funded 'NASA START/TAG or project office liaison' position as part of the mission leadership team.

## 4. SUMMARY AND NEXT STEPS

This brief has presented the Smallsat Technology Accelerated Maturation Platform concept, a dedicated laboratory development and flight demonstration program where accelerated facility development advances technology that feeds a parallel small satellite demonstration mission. This joint facility, laboratory, and prototype instrument development path follows the spirit of the Astro2020 GOMAP concept, as a way of advancing key technologies for all of NASA's Future Great Observatories on a faster track than is possible with the current SAT and small flight mission programs in the ROSES portfolio. Scaling technology while incorporating 'real-world' performance metrics reduces cost and schedule risks to the implementation of large missions – the true purpose of GOMAP. The STAMP combines the three pillars of GOMAP's charge: to advance (1) science definition and (2) technology for the FGOs while (3) providing an inclusive mission framework that develops the science and engineering workforce that will lead the FGOs in the 2030s, 2040s, and beyond.

As an example of the STAMP program, we have described a notional design for "*STAMP-1*": a focused technology, scientific, and workforce building program to advance the technological readiness of ultraviolet hardware for the Habitable Worlds Observatory. Building on the UV technology development roadmap laid out in the LUVOIR Final Report, the cost and development timescale of Pioneers and Explorer Missions of Opportunity, and the short implementation cycle, we estimate *STAMP-1* would require a budget of roughly $50M. Funding for the STAMP program should derive from the FGO/GOMAP line and not pull from or replace existing technology development and small flight opportunities (e.g., the ROSES APRA, SAT, and Pioneers programs).

At the end of the *STAMP-1* mission, the major UV technology and infrastructure maturation efforts for HWO would be complete and team members would ready to support NASA's plans for the full HWO mission implementation. The timeframe for initiating the STAMP program for Habitable Worlds is short – the decadal survey requires a gate review of the progress for Habitable Worlds later in this decade. Therefore, for this project to advance the necessary technology development to a clear path to TRL 5 – 6 ahead of this gate review, NASA should consider developing the necessary programmatic framework to allow STAMP project initiation by 2025, so that the technology and mission development phase for *STAMP-1* can begin in 2025 - 2026.

Code and Data Availability Statement: Data sharing is not applicable to this article, as no new data were created or analyzed.